\documentclass[a4paper,12pt]{article}

\usepackage[backend=bibtex,maxnames=6,bibstyle=numeric,sorting=none]{biblatex}
\usepackage{hyphenat}
\usepackage{authblk}
\usepackage[utf8]{inputenc}  
\DeclareUnicodeCharacter{202F}{\,}
\usepackage[english]{babel}
\usepackage{graphicx}
\usepackage{dcolumn}
\usepackage{bm}

\usepackage{amsmath}
\usepackage{amssymb}
\usepackage{bbm}             
\usepackage{bm}              
\usepackage{color}
\usepackage{latexsym,amsfonts}
\usepackage{graphicx}
\usepackage[                 
	pdfkeywords={meta, information, pdf, hyperref, latex},
        colorlinks=true,
	urlcolor=blue,
	linkcolor=Red2,
        citecolor = magenta,
        pdfpagelabels,
        pdfstartview = FitH,
        bookmarksopen = true,
        bookmarksnumbered = true,
        plainpages = true,
        hypertexnames = false]{hyperref}

\providecommand{\href}[2]{#2}   
\definecolor{Blue2}{rgb}{0.,0.,0.8125}
\definecolor{Brown3}{rgb}{0.625,0.25,0.}
\definecolor{Cyan4}{rgb}{0.,0.56,0.56}
\definecolor{Green4}{rgb}{0.,0.56,0.}
\definecolor{LtBlue}{rgb}{0.27,0.42,0.52}
\definecolor{Magenta4}{rgb}{0.5625,0.,0.5625}
\definecolor{Red2}{rgb}{0.8125,0.,0.}

\DeclareUnicodeCharacter{2212}{-}

\newcommand{\ist}[1]{\overset{\footnotesize(\ref{#1})}{=}}

\title{Against the point-like nature of the electron}
\author[1]{Manfried Faber}
\affil[1]{Atominstitut, Technische Universit\"at Wien,
Stadionallee 2, 1020 Wien, Austria}

\begin{document}
\maketitle


   
\maketitle
Einstein: “You know, it would be sufficient to really understand the electron”, from Hans Dehmelt in~\cite{Dehmelt:1990zz},
\begin{abstract}
Experts in quantum field theory (QFT) generally answer the question of the ``size of an electron'' with ``point-like.'' On the other hand, QFT recognizes quantum effects, shielding by virtual particles, the so-called polarization cloud, which should describe the size of physical electrons. Scattering experiments with electrons, such as those carried out in high-energy experiments at particle accelerators, should be able to clarify whether physical electrons are really point-like, as claimed by experts and in textbooks. In this article, I show that both the formulas of QFT and the corresponding cross sections are consistent with an extent of the electron of the size of the classical electron radius. The assumption that the relativistic energy of electrons in the high-energy limit consists solely of deformation energy from the extended electron density distribution allows for a simple interpretation of the experimental cross sections. For this reason, I refer to classical models in 1+1 and 3+1 dimensions that have precisely this property. The difference between the terms point-like, structureless, and substructureless is highlighted. The usual objections to the claim that the electron radius is finite and has already been measured in electron scattering experiments are discussed.
\end{abstract}

\section{Introduction}
When I ask particle physicists about the size of the electron, the answer is usualy without hesitation that ´the electron is point-like´. Therefore representations of my model of nonlinear electrodynamics~\cite{Faber:2022zwv} wich based on geometrical ideas makes use of topological stable spatially extended solitons frequently triggers the sharp reaction that such a model is unrealistic because `we know that the electron is point-like`. However, the model described in~\cite{Faber:2022zwv}, in which the central object is a classical electron whose field is purely electromagnetic in nature, whose charge is quantized, and which also solves the 4/3 problem of Maxwell's electrodynamics, which has remained unsolved for over 100 years, but which still needs to be supplemented by quantum phenomena as described by the Schrödinger and Dirac equations~\cite{Faber2022}, seems to convey an appealingly clear and concrete understanding of a substructureless electron with a truly finite size. This should serve as a motivation to further explore this way of modeling especially - as we will see - that the generally accepted wisdom of the ´point-like electron` is not established on a real solid basis. When asked how we know that the electron is point-like, reference is made to the regular publications of the Particle Data Group (PDG) and sometimes to studies by Hans Dehmelt from 1988. Experts such as Martinus Veltmann usually express it more cautiously: "The electron is essentially point-like". But Veltman also begins his lecture at the Nobel Laureate Meeting in Lindau in 2015~\cite{Veltman2015} with the words: ``The electron is a small particle, and its charge, as far as we know, is concentrated in a point''. It is therefore understandable that a query about the size of the electron in Wikipedia~\cite{wikiRadiusElectron} yields: ``According to modern understanding, the electron is a point particle with a point charge and no spatial extent.'' Further consultations with experts in quantum field theory (QFT) may eventually lead to the answer: ``The eletron radius is smaller than $10^{-22}$~m.'', a response referring to a work by Hans Dehmelt from 1988 ``A Single Atomic Particle Forever Floating at Rest in Free Space: New Value for Electron Radius''~\cite{HansDehmelt1988}, in which this value is mentioned.

It seems necessary to me to analyze the conclusions that can be drawn from the Standard Model of particle physics regarding the extent of an electron since even distinguished experts in the field of particle physics seem to avoid elaborating about the finite extent of a physical electron. Excellent textbooks~\cite{HenleyGarcía2007} state in headings that ``Leptons Are Point Particles'', thereby misleading students and the physics-interested public with statements about the point-like nature of electrons. It seems incomprehensible that after years of experiments at internationally renowned large research centers with electron-positron colliders, the size of an electron is still unknown. In the following, I will explain that electron scattering experiments and their analysis according to the Standard Model clearly show what the answer should be: “A stationary electron has a size on the order of the classical electron radius and an internal structure that is very homogeneous and symmetrical. However, according to current knowledge, it has no substructure.”

\section{Dehmelt's experiment}

The work of Hans Dehmelt's group~\cite{VanDyck:1987ay} on the magnetic moment of the electron is certainly very fundamental. However, remarkably, Dehmelt's experiments estimating the size of the electron~\cite{Dehmelt:1990zz,HansDehmelt1988} have not been followed up in the meantime~\cite{Mittleman:1992sg,Dehmelt:1999jh}. There is a simple reason for this. Dehmelt draws conclusions for an electron that is composed of three subparticles. He bases his conclusions on a model that investigates such a substructure of the electron~\cite{Brodsky:1980zm}. Dehmelt compares three particles consisting of three subparticles, the triton, the proton, and the electron, and concludes from his measurements that an electron consisting of three subparticles must be smaller than 10$^{-22}$~m. However, such a substructure of the electron is no longer seriously discussed today. One must agree with Dehmelt's conclusion, but it says nothing about the extent of an electron without substructure.

\section{Cross sections}
By measuring cross sections for the scattering of electrons on hydrogen, McAllister and Hofstadter discovered deviations from Coulomb's law for point charges for the first time in 1956, thereby demonstrating that protons are not point particles but have a structure~\cite{Hofstadter1956}. However, the interpretation of the measurement of cross sections suffers in submicroscopic physics from wave-particle duality, so that it is possible to consider the cross section only as a measure of the probability of interaction between incident radiation and the target object (see e.g.~\cite{wikiRadiusCrossSection}). Regardless of this problem, the electric charge radius $\sqrt{}\langle r_E^2\rangle$ of the proton is well defined, as evidenced by its value given by PDG~\cite{ParticleDataGroup:2024cfk}. The most accurate measurement using muonic hydrogen has an accuracy of approximately 0.05\%. It is therefore rather surprising that the PDG does not provide any information about the charge radius of the electron, which is 1836 times lighter. Only the classical electron radius is given,
\begin{equation}\label{rklass}
r_\mathrm{cl}=\frac{e_0^2}{4\pi\varepsilon_0m_ec_0^2}=\alpha_f\frac{\hbar}{m_ec_0}
=2.817 940 3262(13)~\textrm{fm}\quad\textrm{with}\quad
\alpha_f=\frac{e_0^2}{4\pi\varepsilon_0\hbar c_0}.
\end{equation}
In specialized literature on theoretical physics, this radius is often given in “natural” units (nu), with $\hbar=1$ and $c_0=1$, i.e. as
\begin{equation}\label{rklassNat}
r_\mathrm{cl}=\alpha_f/m_e\Big|_\mathrm{nu}.
\end{equation}
This notation in natural units may obscure the fact that $\alpha_f/m_e$ is the natural scale for the extent of the energy density of the electric field of a stationary classical electron with mass $m_e$. The integral over the energy density of a point charge outside $r_\mathrm{cl}/2$ yields  $m_ec_0^2$ as the total energy. From this, one can conclude that the charge density of an extended electron must be maximal in the range of $r_\mathrm{cl}$ in order not to exceed the rest energy of an electron.

The electron-positron experiments at LEP prove that electrons deviate from Coulomb's law for a point-like electron. In accordance with the calculations of quantum field theory (QFT), the fine structure constant $\alpha_f$ at the Z pole~\cite{PDGZPol} is given as 1/127.9. Determining the scale dependence, or “running,” of the coupling constant is a major topic in QFT textbooks. Here, a few sentences on the QFT approach and its interpretation seem necessary.

Electrons and their fields are described by different degrees of freedom, electrons by four-component Dirac spinors $\psi(x)$ and electromagnetic fields by vector fields $\mathcal A(x)$. It must be emphasized here that the fields $\psi(x)$ of the electrons have no further spatial structure. Their interaction is local, meaning that only fields at the same point $x$ in space-time interact with each other through an expression of the form $e_0\bar\psi(x)\mathcal A(x)\psi(x)$, whereby at the point of interaction with the photon field $\mathcal A(x)$, an electron $\psi(x)$ enters and an electron $\bar\psi(x)$ emerges, which can be easily represented in the Feynman diagrams depicting this point-like interaction as a vertex, the meeting point of the two electron lines and one photon line. The charge $e_0$ determines the strength of this interaction. By having three lines meet at the vertices, diagrams with any number of loops can be drawn. The inner lines of Feynman diagrams are referred to as virtual particles because they violate the energy-momentum relations of real electrons and photons. An expansion based on the number of loops and thus on the number of pairs of interaction points in the diagrams proves to be useful due to the small value of the fine structure constant $\alpha_f\propto e_0^2$. The Coulomb interaction is described in QFT by photon exchange, by photon lines between the vertices. As massless particles, photons can travel arbitrarily large or small distances. Fourier transformation of the coordinates to momenta shows that fields with small momenta describe the infinite range of the Coulomb interaction, while arbitrarily high momenta occurring in loops describe the infinitely strong fields at the location of the point-like electrons. The infinities in the algebraic expressions, which arise from the infinitely strong fields, can be compensated for by modifying the fields and coupling constants by momentum-dependent infinite factors, by renormalization, resulting in finite field strengths near the center of the electrons. According to Gauss's law, this procedure corresponds to a smearing of the charge of the electrons. In textbooks, this smearing is vividly illustrated by a polarization cloud of virtual electron-positron pairs, see Fig.\,7.8 in Ref.\,\cite{peskin1995introduction}. Despite point-like and substructure-free mathematical expressions for the fields of electrons and despite local interaction, QFT succeeds in formulating the infinite range of the Coulomb force and a finite extent of the charge distribution of a physical electron. In strong collisions between electrons and/or positrons, such as those that occur in high-energy experiments, this radial dependence of the effective charge distribution manifests itself as an increase in the strength of the Coulomb interaction with the available energy. To clarify this difference between the mathematical formulation and the natural quantities, field theory distinguishes between bare and effective quantities. In this case, a distinction must be made between the point-like charge distribution of the bare electron and the extended effective charge distribution of a physical electron. Scattering experiments should be ideally suited to experimentally determining such effective physical density distributions.

Hofstadter's experiments on electron scattering on protons and atomic nuclei do not reveal much about electron radii, because the evaluation of the experiments assumes that electrons are point-like and therefore have no internal structure. In comparison, electron-electron and electron-positron scattering, or even photon scattering, seem much more meaningful. These scatterings can be calculated precisely in QFT and their results correspond excellently with the experiments. The lack of statements about the size of an electron can therefore only lie in the interpretation of the results already assuming the point-like structure of the electrons.

This scattering of photons by electrons, known as Compton scattering~\cite{dumond1929compton,heitler1954qtr}, is described in QFT by the Klein-Nishina formula~\cite{KleinNishina1929}. The differential cross section for photons with wave numbers $k_i\to k_f$ and polarizations $\vec\varepsilon_i\to\vec\varepsilon_f$~(5-112\cite{ItzyksonZuber1985}) is
\begin{equation}\label{diffThomsonQuerschnitt}
\frac{d\sigma}{d\Omega}=\frac{r_\mathrm{cl}^2}{4}\Big(\frac{k_f}{k_i}\Big)^2
\Big[\frac{k_f}{k_i}+\frac{k_i}{k_f}+4(\vec\varepsilon_i\cdot\vec\varepsilon_f)^2
-2\Big].
\end{equation}
It is interesting that this expression is independent of the energy scale of the photons due to the ratios $\frac{k_f}{k_i}$. This indicates that the value of the cross section is a property of the electron and not of the photons. One might think that the square of the length scale that must appear in a cross section is based solely on dimensional considerations and is not a real radius. In this case, wouldn't the Compton wavelength be more appropriate than the electron radius, since it appears in the well-known, kinematically based Compton scattering formula $\lambda_i-\lambda_f=\hbar/(m_ec_0)(1-\cos\vartheta)$, where $\lambda_{i,f}$ are the wavelengths of the incident and outgoing radiation and $\vartheta$ is the scattering angle? There must therefore be a reason why the classical electron radius $r_\mathrm{cl}$ is written in Gl.~(\ref{diffThomsonQuerschnitt}).  The exact size of the prefactor is easiest to interpret for elastic scattering $k_i=k_f$ while maintaining the polarization direction $\vec\varepsilon_i=\vec\varepsilon_f$
\begin{equation}\label{elastThomsonQuerschnitt}
\frac{d\sigma}{d\Omega}\Big|_{k_i=k_f,\vec\varepsilon_i=\vec\varepsilon_f}
\ist{diffThomsonQuerschnitt}r_\mathrm{cl}^2.
\end{equation}
It corresponds to a process in which photons are absorbed on a surface element $r_\mathrm{cl}^2\,d\Omega$ of a sphere with radius $r_\mathrm{cl}$ and emitted in the same direction and with the same polarization.

The scattering of very low-energy photons is called Thomson scattering. The spin-averaged total Thomson cross section also has a very simple form (5.92\cite{peskin1995introduction})
\begin{equation}\label{ThomsonQuerschnitt}
\sigma_\mathrm{T}=\frac{8}{3}\pi r_\mathrm{cl}^2
\end{equation}
and thus differs very little from the naively expected value $\pi r_\mathrm{cl}^2$. The effective cross section of an electron, which has no sharp edge due to the shape of its electric field, can be compared to that of a hard sphere, resulting in a radius of approximately $1.6\,r_\mathrm{cl}$. The slightly larger value than $r_\mathrm{cl}$ is easy to understand when you consider that in the calculation of $\sigma_\mathrm{T}$ from Eq.~(\ref{elastThomsonQuerschnitt}), quantum mechanics is used to sum over the possible exit channels~\cite{ItzyksonZuber1985}. Due to the accuracy of the measurements, $\sigma_\mathrm{T}$ is no longer determined experimentally, but is calculated from the fundamental constants~\cite{nist_thomson_2022}. Since the value~(\ref{ThomsonQuerschnitt}) for $\sigma_\mathrm{T}$ agrees very well with the experiment, the question inevitably arises as to how it can be that in scattering experiments in high-energy accelerators, the electron can be regarded as point-like. To answer this question, one must take into account that since Hofstadter, the energies of electrons have increased 1000-fold. The Lorentz contraction that occurs at these high energies is explicitly stated in the expressions given for the cross sections, but it is not really mentioned in the analysis of the scattering experiments.

To facilitate the explanation of high-energy experiments and their description by QFT, I will study classical models on the following pages that make important aspects of these results understandable. I will present surprisingly simple Lorentz-covariant field models in 1+1 dimensions and in 3+1D, which are characterized by topological excitations with particle character and relativistic behavior, whose mass consists only of spatially highly concentrated field energy. These extended particles are characterized by topological properties such as winding numbers and are therefore referred to as topological solitons. Such models explain the structure of particles and their mutual forces through geometrization. In the high-energy limit, when the rest mass of these particles is negligible, the relativistic mass increase can be explained solely by the internal stress energy, which increases proportionally to the Lorentz contraction, i.e., to the Lorentz factor $\gamma$. In a highly relativistic central collision between repelling particles, the relativistic mass is completely converted into deformation energy at the reversal point. This important property is a prerequisite for the thickness of the particles to decrease by a factor of $\gamma$ in strong collisions. In 3+1D, spherically symmetric density distributions are compressed into ellipsoids, whose shape is determined by Lorentz contraction in the direction of flight and by impact forces in the direction of the relative coordinate. This compression occurs for a given direction of flight for each orientation of the scattering plane, thus explaining a decrease in the cross section by a factor of $\gamma^2$.

For demonstration I will start with a very illustrative purely mechanical experiment which surprisingly even demonstrates the effects of Lorentz contraction in 1+1D, in which a series of coupled pendulums are attached at equal intervals along a horizontal, suspended piano string, as described in Chapter 6 of Remoissenet's book, see Fig.\,6.5 in Ref.\,\cite{remoissenet:1999wa}. Assuming frictionless motion of the pendulums, the so called Sine-Gordon model in natural coordinates of position $x$ and time $t$ provides an analytical description of the orientation $\theta$ of the pendulums through the Lorentz-invariant differential equation (6.10\cite{remoissenet:1999wa})
\begin{equation}\label{SGDG}
\partial_t^2\theta-\partial_x^2\theta+\sin\theta=0.
\end{equation}
In these natural coordinates, the signal propagation velocity has a value of one. Of particular interest are the kink and antikink solutions~(6.12\cite{remoissenet:1999wa}),
\begin{equation}\label{loesungen}
\theta(x,t)=4\,\arctan\big[\exp\big(\pm\gamma(x-x_0-\beta t)\big)\big]\quad
\textrm{with}\quad\gamma:=\frac{1}{\sqrt{1-\beta^2}},
\end{equation}
stable topological excitations in which the orientation of the pendulums describe a topological soliton, a helical transition between two neighboring vertical orientations $\theta=2\pi n,\;n\in\mathcal Z$ favored by gravity. The distance $x_+-x_-=1.76$ between horizontal orientations $\theta(x_-,0):=\pi/2$ and $\theta(x_+,0):=3\pi/2$ of static solutions is shortened in moving solutions by the well-known Lorentz factor $\gamma$. It is interesting to analyze how this Lorentz contraction occurs in mechanical pendulums: When a soliton moves, the more strongly moving pendulums must signal to the less moving pendulums that they must move faster. This is only possible by increasing the angle between neighboring pendulums, so that the size of a soliton is necessarily reduced and the energy stored in the soliton increases exactly by the $\gamma$ factor~(6.24\cite{remoissenet:1999wa}).

\begin{figure}[h!]
\centering
\includegraphics[scale=1.0]{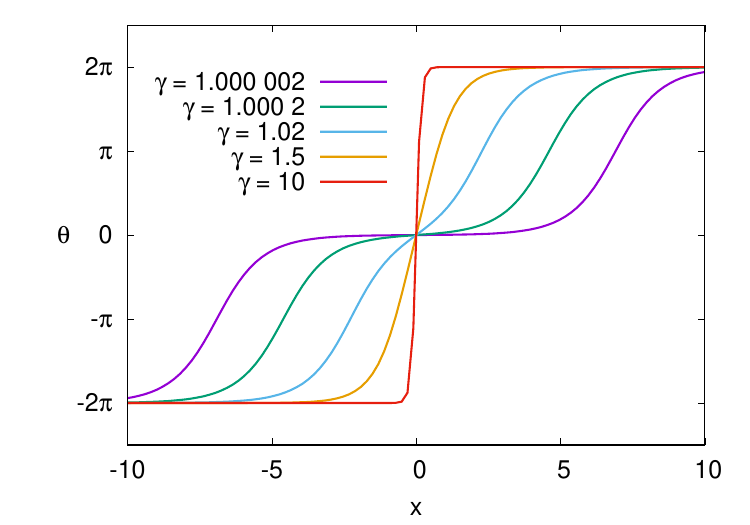}
\caption{Profiles $\theta(x,t=0)$ for the colliding solitons of Eq.~(\ref{ZweiSolitonTheta}) at the reversal point for different velocities $\beta$ and corresponding Lorentz factors $\gamma$. Although slow-moving solitons ($\gamma\approx 1$) have a finite size, they can come arbitrarily close to each other when they collide at high relative velocities.}
\label{eekollision}
\end{figure}

The time-dependent two-soliton solutions and soliton-antisoliton solutions are also known analytically, see (6.29\cite{remoissenet:1999wa}) and (6.30\cite{remoissenet:1999wa}). They are very useful for investigating the effects of high velocities. When two solitons of the same kind collide at high speed, they cannot penetrate each other for topological reasons, but they can approach each other to arbitrarily small distances, as can be seen from the soliton-soliton solutions~(6.29\cite{remoissenet:1999wa})
\begin{equation}\label{ZweiSolitonTheta}
\theta(x,t)=4\,\arctan\frac{\beta\,\mathrm{sinh}(\gamma x)}{\mathrm{cosh}(\beta\gamma t)},
\end{equation}
where $\pm\beta$ is the velocity at which the solitons approach from infinity. At time $t=0$, they come to rest and then move in the opposite direction. The profiles for $t=0$, the moment of closest approach, are shown in Fig.~\ref{eekollision}.
Although stationary solitons have a finite extent of 1.76 natural length units, their homogeneous internal structure allows them to approach each other at any distance. The contraction described by the Lorentz factor at large distances is not lost during the collision, but is converted into a contraction due to the strong braking forces. It is easy to imagine what the profile function of the collision of two solitons should look like for $\gamma=2\cdot10^5$.

When a macroscopic elastic body is thrown against a wall, its Lorentz contraction is minimal, and the deformation upon impact is much greater. This raises the question of why, in the Sine-Gordon model, the deformation at the reversal point assumes the value of the Lorentz contraction at high speeds. For $\gamma\to\infty$, the potential energy~(6B.6\cite{remoissenet:1999wa}) disappears and the kinetic energy~(6B.6\cite{remoissenet:1999wa}) is zero at the reversal point. The stress energy~(6B.4\cite{remoissenet:1999wa}), which is proportional to $\gamma$, remains as the only energy contribution~(13\cite{faber43-2025}). It is therefore understandable that it must absorb all of the kinetic energy, which is proportional to $\gamma$ when the rest energy is negligible.

\begin{figure}[h!]
\centering
\includegraphics[scale=1.0]{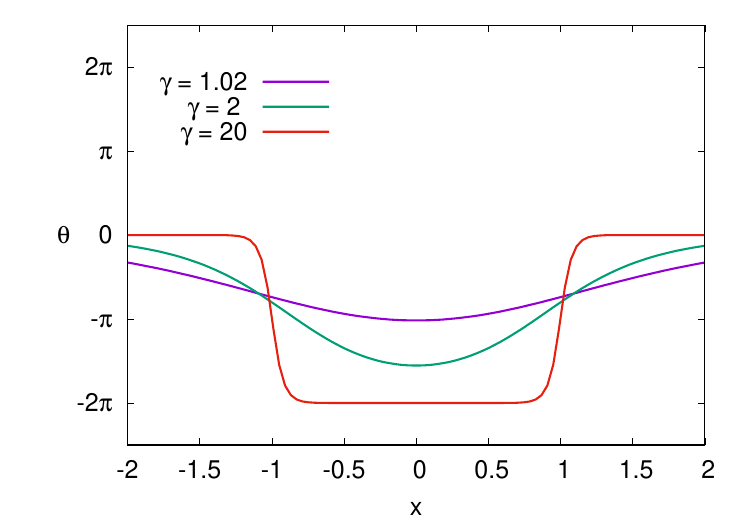}
\caption{Rotation angle $\theta(x,t=-1)$ in the kink-antikink system of Eq.~(\ref{SolitonAntisolitonTheta}) for different values of $\gamma$, if an antikink enters from $x=-\infty$ and a kink from $x=+\infty$.}
\label{pevorher}
\end{figure}

In the kink-antikink system, the attraction between particles and antiparticles has a geometric origin and also explains their annihilation in an interesting way. In the configuration proposed in (6.30\cite{remoissenet:1999wa}),
\begin{equation}\label{SolitonAntisolitonTheta}
\theta(x,t)=4\,\arctan\frac{\mathrm{sinh}(\beta\gamma t)}{\beta\,\mathrm{cosh}(\gamma x)},
\end{equation}
an antikink comes from $x=-\infty$ and a kink from $x=+\infty$. For sufficiently negative times, the region between kink and anti-kink has a negative angle of $-2\pi$, as shown in Fig.\,\ref{pevorher}. Annihilation begins slightly before $t=0$ and is complete at $t=0$, with the angle disappearing identically. At this point in time, all of the energy is contained in the rapid change of the field. The speed of the zero crossing is at its maximum at $x=0$. For very high collision energies, this speed assumes equally high values and lasts only for an extremely short time. At positive times, a particle-antiparticle pair is created again and the angle has changed its sign to positive values and develops between kink and antikink towards $\theta=2\pi$.

We can now apply the insights gained from the Sine-Gordon model to electron-electron and electron-positron scattering and make predictions, which we then compare with the experiments and the predictions of QFT. We start from a generalization of the Sine-Gordon model from 1+1 D in space and time to 3+1\,D, as first presented in~\cite{Faber:1999ia}. In this generalization, it is necessary to generalize the angle $\theta$ of the Sine-Gordon model to three angles, e.g., the three Euler angles or the rotational vector, see Eq.~(2\cite{Faber:2022zwv}). In this model, the electron is described by a stable spherically symmetric soliton. The energy density~(33\cite{Faber:2022zwv}) of the corresponding purely electromagnetic field is mainly concentrated in a spherical volume with radius $r_\mathrm{cl}$.

For electron-electron scattering in the center-of-mass system, we expect such extended solitons to be Lorentz-contracted in the direction of flight, i.e., to look like a pancake or a beer mat~\cite{Faber:1999ia}, as we know from heavy ion experiments at the LHC. At the highest energies available at LEP2, the Lorentz factor was approximately $\gamma=2\cdot10^5$. This corresponds to a 'thickness\,:\,circle radius' ratio of a circular sheet of paper with a radius of 20 m or a beer mat with a radius of 200 m. When two such flat solitons collide, they turn their flat sides toward each other in order to distribute the repulsion evenly. They are slowed down by mutual repulsion, and the contraction is maintained by the braking effect, analogous to the behavior of the kink-kink solution in the Sine-Gordon model, see Fig.~\ref{eekollision}. Since, in the highly relativistic limit, only the deformation energy contributes to the mass at the turning point, this behavior is to be expected. The particles can therefore approach each other much closer than the radius corresponding to their rest state.

During collisions, electrons are subject to three influences that affect their behavior~\cite{Faber:1999ia}:
\begin{itemize}
\item Lorentz contraction in the direction of flight.
\item Forces exerted by the collision partners in the direction of the relative distance.
\item A homogeneous geometric internal structure that manifests itself in the stability of stationary particles and can hold electrons together despite collisions, unless there is sufficient energy to generate a shower of particles.
\end{itemize}
\begin{figure}[h!]
\hspace{-30mm}\begin{minipage}[c]{45mm}
\includegraphics[scale=1.0]{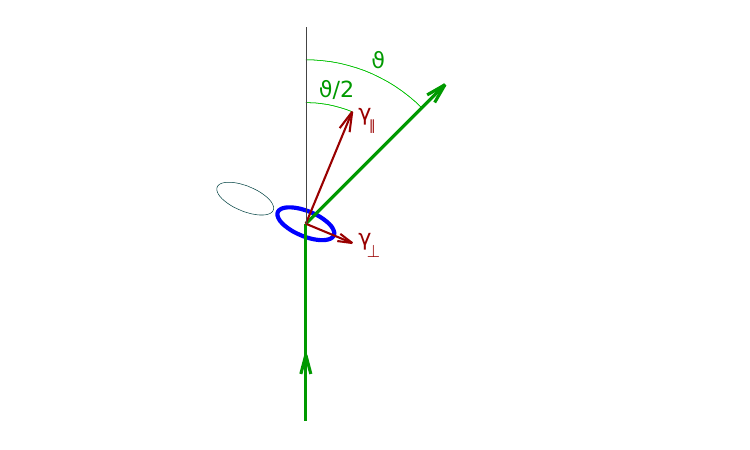}
\end{minipage}
\begin{minipage}[c]{45mm}
\includegraphics[scale=1.0]{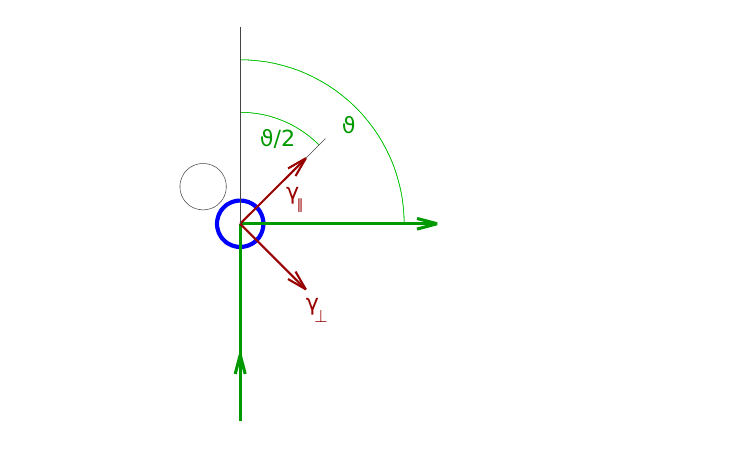}
\end{minipage}
\begin{minipage}[c]{40mm}
\hspace{-20mm}\includegraphics[scale=1.0]{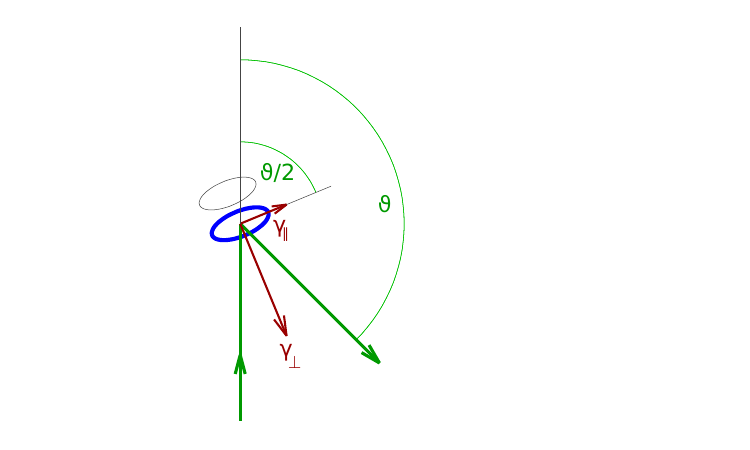}
\end{minipage}
\caption{Electron-electron scattering is represented in the scattering plane for scattering angles of 45°, 90°, and 135° by a thick green line at the point of closest approach, the periapsis. The density distribution of each electron is distorted by Lorentz contraction $\gamma_\parallel$ in the direction of flight $\vartheta/2$ and by the collision component $\gamma_\perp$ in the orthogonal direction, and is shown as a blue ellipse. The respective position of the collision partners is symbolized by a thin gray line.}
\label{streuung}
\end{figure}
In highly relativistic collisions $\gamma\gg1$, the electrons do not follow a Keplerian hyperbola, as would correspond to a spherically symmetric $1/r$ potential, but are strongly repelled over a short distance of $\gamma r_\mathrm{cl}$. The deformation force responsible for this is strongest at the point of closest approach, the periapsis. The shape and position of the extended particles at the periapsis therefore provide the best information about the extent of the scattering bodies; see Fig.\,\ref{streuung} for three characteristic scattering angles. At the periapsis, the electrons have only completed half of their change in direction $\vartheta$ and are moving in the direction $\vartheta/2$ at a reduced speed, as corresponds to the Lorentz factor
\begin{equation}\label{GammaParallel}
\gamma_\parallel=\gamma\cos\frac{\vartheta}{2}.
\end{equation}
The deformation force acts in the scattering plane orthogonally to the direction of flight, i.e., in the direction $(\vartheta+\pi)/2$, and is proportional to $\gamma_\perp=\gamma\sin\frac{\vartheta}{2}$.

For scattering angle $\vartheta=\pi$, i.e., for central collision of the two disks, the interaction at the reversal point extends over the full width of the ellipsoids for purely geometric reasons. The velocity in the tangential direction of the interaction zone is zero. As the scattering angle decreases, the velocity and consequently the Lorentz contraction in the tangential direction increase according to Eq.~(\ref{GammaParallel}), and the width of the interaction zone therefore decreases until it reaches the thickness of the flat ellipsoid at the peripheral collision.

Since we are considering models whose relativistic mass increase results from the increase in stress energy, $\gamma_\perp=\gamma$ applies to central collisions. As the scattering angle decreases, the collision becomes weaker and the compression $\gamma_\perp$ due to pressure becomes continuously smaller. For a scattering angle $\vartheta=\pi/2$, Lorentz contraction and compression due to the collision are equal in magnitude. The cross-sectional area of the electron in the scattering plane is then most circular, as shown in the middle diagram in Fig.\,\ref{streuung}. However, normal to the scattering plane, the electron always has the full extent of the rest electron.

The angle of rotation, which is shown in Fig.\,\ref{streuung} at the periapses by the deviation of the ellipses from orthogonality to the direction of incidence, is somewhat more complicated. Due to the opposite movement, the two discs fly parallel toward each other and parallel away from each other at large distances for all scattering angles. Therefore, for scattering angles $0<\vartheta<\pi$, the discs must rotate during the collision period. For $\vartheta=\pi$, there is no rotation of the ellipsoids.  Up to $\vartheta=\pi/2$, the rotation increases to $\pi/2$ in accordance with the scattering angle. For the scattering angle $\vartheta=\pi/2$, due to the circular cross-section at the periapsis, the rotation can take place through a change in shape, with the major axis of the ellipse jumping by 90°. As the scattering angle continues to decrease, the rotation decreases to zero. Increasingly thinner edge areas of the ellipsoids collide with each other, as illustrated by the left diagram in Fig.\,\ref{streuung}.

We expect the contraction described above with the factor $\gamma$ for any orientation of the scattering plane, i.e., in the two dimensions normal to the direction of impact. This leads to a decrease in the cross section with the square of the Lorentz factor, with  $\gamma^2$,
\begin{equation}\label{Vorhersage}
\sigma_\mathrm{predicted}=\frac{r_\mathrm{cl}^2\pi}{\gamma^2}
\end{equation}
Two compensating effects had to be taken into account for the radius of this circular area. On the one hand, only the reduced masses contribute in the laboratory system of a collider; on the other hand, the expansion of both collision partners contributes to the interaction area.

There are no decelerating effects for electron-positron scattering; the particles are actually accelerated toward each other by Coulomb attraction, so that Lorentz contraction tends to increase. The resulting flattened ellipsoids face each other with their flat sides in order to distribute the attraction evenly. Annihilation can essentially only occur if the collision parameter is of the order of magnitude of the Lorentz-contracted classical electron radius $r_\mathrm{cl}/\gamma$. At larger distances, the two disc-shaped ellipsoids slide past each other undisturbed. Again, a cross section similar to that for electron-electron collisions, i.e., Eq.~(\ref{Vorhersage}), is to be expected.

The preceding considerations, and in particular Eq.\,(\ref{Vorhersage}), have clearly shown that it is only necessary to assume a finite size of electrons of the order of $r_\mathrm{cl}$ in order to be able to make many important predictions about cross sections in electron scattering experiments. In addition to these purely classical effects, I expect quantum effects due to wave-particle duality. A description of the particle trajectories in Feynman's path integral formalism shows that only trajectories whose action deviates from the extremal trajectory by a few $\hbar$ contribute to the transition amplitude. Due to the high Lorentz factor mentioned above, only particle trajectories that come very close to the classical trajectories need to be considered. Only in the region of closest approximation should the quantum mechanical transition amplitudes play a role for a brief moment. This suggests a model reminiscent of the compound nucleus model~\cite{Lilley2001,Kamal2014} in nuclear physics. The transition amplitudes should be a product of three factors: contributions from the input channel $A_\mathrm{in}$, the region of closest approximation $A_\mathrm{ca}$, and the output channel $A_\mathrm{out}$
\begin{equation}\label{DreifachProdukt}
A=A_\mathrm{in}A_\mathrm{ca}A_\mathrm{out}.
\end{equation}
While $A_\mathrm{in}$ and $A_\mathrm{out}$ can essentially be calculated using classical methods, quantum mechanical treatment is essential for calculating $A_\mathrm{ca}$ in order to understand the contributions of the many output channels of high-energy scattering experiments and their energy dependence.

It stands to reason that the clearest experimental information about the structure of an electron can be obtained besides photon scattering from e$^-$e$^-$ and e$^-$e$^+$ collisions. For both types of collision experiments, there are many outgoing channels, which usually have to be treated separately in QFT. As far as the size of the electron is concerned, all respective processes contribute to the area that electrons oppose to the collision partners. Purely electromagnetic processes are the M{\o}ller scattering~\cite{Moller1932}, e$^-$e$^-$\;$\to$\;e$^-$e$^-$, Bhabha scattering~\cite{Bhabha1936}, e$^-$e$^+$\;$\to$\;e$^-$e$^+$, and annihilation e$^-$e$^+$\;$\to$\;$\gamma\gamma$. Many chapters in good textbooks on QFT are devoted to the calculation of cross sections~\cite{peskin1995introduction,ItzyksonZuber1985,kaku1993qft}. The final results and their limiting cases are given in more or less detail. Due to the complexity of the quantum field theory treatment, I will therefore only make some basic remarks here. Ultimately, the only question is whether electrons should be regarded as extended objects. This question can be answered solely by interpreting the total cross sections. Details such as the angular dependence of the cross section for spherically symmetric density distributions can be derived mainly from kinematics. A great deal of insight has been gained from the resonances that appear in the energy dependence of the cross sections. These concern complex questions of quantum field theory, but are not relevant to the fundamental consideration of the size of stationary electrons and their shape change in high-energy experiments. Since the arguments for the point-like nature of the electron are mainly supported by high-energy experiments, it is sufficient to study the high-energy behavior of the processes.

All processes have in common that the QFT calculations of the cross sections in natural units (nu) in the high-energy limit, where the particle masses can be neglected, contain the factor
\begin{equation}\label{gemeinsam}
\frac{\alpha_f^2}{s}\Big|_\mathrm{nu}=\frac{\alpha_f^2}{4m_e^2\gamma^2}\Big|_\mathrm{nu}
=\frac{r_\mathrm{cl}^2}{4\gamma^2},
\end{equation}
where $s$ is the square of the center-of-mass energy in the laboratory system, i.e., $s=(2\gamma m_ec_0^2)^2$. The differential cross sections for M{\o}ller scattering are given in (6.42-6.43\cite{ItzyksonZuber1985},6.49-6.51\cite{kaku1993qft}) and for Bhabha scattering in (6.47-6.49\cite {ItzyksonZuber1985}, 6.54-6.56\cite{kaku1993qft}). Because the range of the electric interaction extends to infinity, the total cross section for these processes diverges. Due to the good theoretical understanding of Bhabha scattering, its small-angle scattering is used to determine the luminosity, i.e., the number of particles per unit area and time. For electron-positron scattering with a muon pair in the output channel, the differential cross section is finite and gives~\cite{peskin1995introduction}
\begin{equation}\label{eemumu}
\sigma_\mathrm{tot}=\frac{4\alpha_f^2}{3s}\Big|_\mathrm{nu}
=\frac{r_\mathrm{cl}^2\pi}{3\gamma^2}.
\end{equation}
This process is used as a reference process for the production of quark-antiquark pairs and subsequent hadronization, in which the number of quark colors plays an important role alongside the number and charge of the quark flavors.

The factor\,(\ref{gemeinsam}), which is essential for the order of magnitude of the QFT cross sections, emphasizes the importance of the classical electron radius $r_\mathrm{cl}$ for the size of the polarization cloud around electrons and underpins the validity of the classical considerations expressed by Eq.\,(\ref{Vorhersage}). Since the Thomson cross section (\ref{elastThomsonQuerschnitt}) calculated according to QFT also agrees very well with $r_\mathrm{cl}$, it would be interesting to know whether QFT calculations of the energy density around electrons also predict a concentration of the rest energy of an electron around $r_\mathrm{cl}$.

\section{Conclusion}
The predictions of quantum field theory (QFT) regarding the cross sections in collision experiments are excellent. Many insights have been gained from the analysis of these experiments. In this work, I attempt to show that the expressions derived from QFT are compatible with the interpretation that stationary electrons have a spherically symmetric density distribution with an extension on the order of the classical electron radius. Outside this central area, the energy density decreases with the square of the electric field strength, i.e., with the fourth power of the radial distance $r$. In high-energy experiments, these density distributions are Lorentz-contracted and take on an ellipsoidal shape in the laboratory system. Since the Lorentz factor reaches a value of $2\cdot10^5$ at the highest energies, these ellipsoids are very flat and slide past each other during scattering. This can give the impression that electrons are point-like. To date, there is no evidence that electrons have a substructure, i.e., that they are composed of subparticles.

This leaves the question why the view that ``electrons are essentially point-like objects'' is more or less unquestioned accepted up to this day. To understand this we should look at the history of the development of QFT. Since the discovery of electrons, there has been no model capable of describing the stability of an electron. This is known in the history of physics as the 4/3 problem~\cite{faber43-2025}. In the early decades of QFT, the electron was therefore described as a point-like particle, which led to infinities in the calculations. Veltman describes the situation very vividly in one of his lectures~\cite{Veltman2015}: {\em'The general feeling was we have to do something about quantum mechanics. But what? The trouble is, while you were busy with field theory, as created by Pauli and Heisenberg, you got infinities all over the place. So, it's very, very demotivating if you're busy calculating and you cannot go ahead because there's an infinity coming up; you know nature is finite. But then there came Mr. Kramers, a professor in Leiden, he came with a terrific idea, and that was the idea of renormalization. He realized something that people had not realized before. It was a very simple idea, like there are so many simple ideas that you never know, and that goes about the electron self-energy. You see, the electron, we picture it as something which has electric charge, and the way we think about it is, the electron has self-energy. That is to say it's the electric charge that you press together that you call the self-energy of the electron. Now, the electron may also have energy of its own, it has a mass, so it could have a certain energy. What happens is that the mass of the electron that you observe finally, is the result of two things, the addition of two things, namely, the mass that you start off plus the self-energy that is generated by the electric charge. So, the experimental mass, such as we see it, is the bare mass, that is the mass of the electron that somewhere has been chosen by God to be there and then in addition, you get the energy due to all these charges being compressed. The sum of these two is what you experimentally observe. So, Mr. Kramers said: ``Well, let's do it like follows: Let's not try to understand everything. Let's just say the experimental mass, that's something we can measure and God knows what goes on in the electron at small distances and the like. Why don't we just skip that part of the problem?'' So, what Mr. Kramers did was, he said: ``Okay, the self-energy of the electron is infinite. Okay, what we will do is, we will start off with an electron with the opposite amount of self-energy and so the only thing we can say is the sum of the two should be what we see experimentally.'''}. This proposal was very successful and, as mentioned, led to many new insights, so there was no pressure to solve the problem that Kramer had overcome. Instead, the opinion developed that the problem had been solved anyway. This however is a purely mathematical solution which possibly hides what is really going on at small distances.

Following Veltman's historical remarks, I would like to address an even more fundamental question: To what extent can mathematics contribute to our understanding of nature? Galileo Galilei undoubtedly opened humanity's eyes with his statement  ``La filosofia è scritta in questo grandissimo libro che continuamente ci sta aperto innanzi a gli occhi (io dico l’universo), ma non si può intendere se prima non s’impara a intender la lingua, e conoscer i caratteri ne’ quali è scritto. Egli è scritto in lingua matematica$\dots$''~\cite{galilei1623saggiatore}\footnote{Philosophy is written in that great book which is ever open before our eyes (I mean the universe), but it cannot be understood unless one first learns to understand the language and to know the characters in which it is written. It is written in the language of mathematics\dots}. But isn't Arthur Schopenhauer also right when he writes~\cite{Schopenhauer}: ''Every understanding is an immediate one, and therefore an intuitive apprehension of the causal connection, although it must readily become changed into abstract notions so as to become established.'' Shouldn't there first be an intuitive understanding, the accuracy of which should be verified by mathematical methods? QFT has taken the opposite approach. It has chosen a mathematical method, applied it consistently, and drawn interesting conclusions, such as the dependence of the magnitude of the electric charge on the strength of the collisions, known as “running coupling.” However, this method has the disadvantage that it runs the risk of going up the wrong path, which may lead to a dead end where it is no longer possible to make real progress because the explanations have to become increasingly complicated, just as when we climb a mountain from a valley that becomes narrower and narrower and we eventually reach dangerous terrain. Then it would certainly be wiser to turn back. In my opinion, it would be wise to look for phenomena in nature that are analogous to the process we want to investigate. These can lead to an intuitive understanding. If we have already incorporated a phenomenon into our worldview, we can also intuitively understand the analogous problem. With the help of mathematics, we can then try to describe this phenomenon without contradiction. Mathematics should be the tool that verifies the correctness of logical conclusions.  This means nothing other than our strong belief that nature follows laws that we can rely on. That is precisely what we experience every day.

I would like to add another comment here regarding the possible realization of an extended, purely electromagnetic electron and its high-energy behavior. A finite size of electrons means that the energy density inside the electron does not have to be infinite, but can tend toward a constant value. The radial dependence of the field strength could, as Schwinger~\cite{Schwinger1983} suggested, decrease with the radial distance $r$ as $1/(r^2+r_0^2)$, which would lead to a finite total energy of the electric field. However, this field distribution would not be stable because an increase in $r_0$ would lead to a decrease in total energy, causing such an electron to explode. This manifests itself in the 4/3 problem of the classical electron. Its precise treatment shows~\cite{faber43-2025} that internal stresses would exist in this electron. These stresses can be eliminated by allowing non-Abelian fields inside a stable electron. In order to stabilize such a non-Abelian field configuration, which transitions into an Abelian field configuration at infinity, a compressive contribution to the energy is required, analogous to the weight of the pendulums in the pendulum model mentioned above. Pendulums that point upwards need additional energy and increase the total energy. A balance is needed between the compressive and smoothing forces, which in such problems is expressed in a virial theorem, an optimal ratio between the two opposing tendencies. In the aforementioned extension of the Sine-Gordon model to 3+1 dimensions, the virial theorem states that the potential energy must be one quarter of the total energy, see Eq.~(23\cite{Faber:2022zwv}). If we add the 25\% share of potential energy to the three quarters of electromagnetic energy, the rest energy increases by a factor of 4/3, thereby solving the 4/3 problem~\cite{faber43-2025} of classical Maxwellian electrodynamics, which does not recognize any contribution to potential energy. If the electron field is purely electromagnetic in nature, as Abraham~\cite{Abraham1903} and Lorentz~\cite{Lorentz1909} assumed, then in the limiting case $\gamma\to\infty$, the contributions of the potential energy~(59\cite{Faber:1999ia}) and the parallel electric and magnetic fields~(57\cite{Faber:1999ia}) disappear. Thus, only the two transverse electric field components and the two transverse magnetic field components contribute to the energy~(A47\cite{faber43-2025}). All four are proportional to $\gamma$~(A49\cite{faber43-2025}), which explains why, as mentioned above, the contraction with the Lorentz factor $\gamma$ is maintained in high-energy electron-electron collisions despite the deceleration.

Finally, a comparison with the Sine-Gordon model and its 3+1-dimensio\-nal generalization leads to another interesting conclusion. Both models are topological field models characterized by their topological field quantization. A field quantization via harmonic oscillators thus becomes unnecessary. This eliminates the need for field quantization by harmonic oscillators. This quantisation may be so successful because the harmonic oscillator spectrum is equidistant and thus analogous to the spectrum of free particles. Topological field quantization therefore avoids the problem that QFT has with explaining the cosmological constant, which, according to Weinberg~\cite{Weinberg:2000yb}, makes a prediction for the cosmological constant that is incorrect by 120 orders of magnitude, because the zero-point energies of the normal modes of all fields must be added together.

In summary, it can be said that all experiments performed with electrons to date indicate that electrons have no substructure. In this article, I have presented arguments in favor of a finite size of electrons being compatible with QFT calculations, so these considerations do not cast doubt on the accuracy of the mathematical calculations. However, a finite extent of an electron leads to an almost constant finite energy density at the center and would therefore have far-reaching consequences for our understanding of physics. It is understandable that such a proposal is met with a great deal of resistance. We humans are obviously easily tempted to reject anything that does not fit into our worldview.

\section*{Acknowledgements}
I would like to express my sincere thanks to Manfred Jeitler, Jeff Greensite and Rudolf Golubich for their critical comments, Johannes Werner and Martin Suda for their help in making the text more understandable, and Jaroslaw Duda for his impressive reference to Fig.\,46.6\,(top) in Ref.\,\cite{Beringer2012PDG} for the total cross sections $e^+e^-\to\,$hadrons, whose continuation to $\sqrt{s}=1$ MeV leads to the cross section of a classical electron.

\printbibliography[title={References}]
\end{document}